\documentclass[superscriptaddress,twocolumn,prl]{revtex4}

\usepackage{graphicx}
\usepackage{amsmath}
\usepackage{longtable}
\usepackage{amssymb}
\usepackage{latexsym}
\usepackage{amsmath}
\usepackage{natbib}

\begin{document}

\title{Phase-modulated decoupling and error suppression in qubit-oscillator systems}

\author{Todd J. Green}
\author{Michael J. Biercuk}
\email[To whom correspondence should be addressed: ]{michael.biercuk@sydney.edu.au}

\affiliation{ARC Centre for Engineered Quantum Systems, School of Physics, The University of Sydney, NSW 2006 Australia}

\date{\today}

\begin{abstract}
We present a scheme designed to suppress the dominant source of infidelity in entangling gates between quantum systems coupled through intermediate bosonic oscillator modes.  Such systems are particularly susceptible to residual qubit-oscillator entanglement at the conclusion of a gate period which reduces the fidelity of the target entangling operation.  We demonstrate how the exclusive use of discrete phase shifts in the field moderating the qubit-oscillator interaction - easily implemented with modern synthesizers - is sufficient to both ensure multiple oscillator modes are decoupled and to suppress the effects of fluctuations in the driving field.  This approach is amenable to a wide variety of technical implementations including geometric phase gates in superconducting qubits and the Molmer-Sorensen gate for trapped ions.  We present detailed example protocols tailored to trapped-ion experiments and demonstrate that our approach allows multiqubit gate implementation with a significant reduction in technical complexity relative to previously demonstrated protocols.

\end{abstract}


\maketitle
Quantum mechanical entanglement is an important resource for a new generation of quantum-enabled technologies, most notably quantum information processing (QIP) \cite{NC2010}.  A key requirement for scalable QIP is the ability to controllably produce high-fidelity multi-particle entanglement on demand.  This is accomplished in experimental systems using a variety of techniques, but a prominent approach relies on the realization of an indirect interaction between basic quantum systems (here qubits) mediated by bosonic oscillator modes \cite{Monroe1995,Sorensen1999,Milburn2000,Feng2007,Schoelkopf2007,Albrecht2013}. A significant source of infidelity in these experiments is the presence of residual qubit-oscillator entanglement at the conclusion of an interaction period, leading to decoherence and a degradation of the fidelity of entanglement generation. Therefore, the ability to effectively and efficiently disentangle qubits from bosonic modes is vital for many modern experimental implementations of entangling QIP operations.

In this letter, we describe a simple technique to \emph{decouple} qubits from multiple intermediary bosonic modes in order to improve entangling-gate fidelity.  The technique is based solely on technologically simple, discrete shifts in the \emph{phase} of the driving field that mediates the qubit-oscillator coupling. We present a generalized theoretical framework permitting construction of protocols providing suppression of residual qubit-oscillator couplings in a densely packed mode structure.  In addition to ensuring that all excited modes are decoupled under the assumption of a perfect driving field, we demonstrate how the same framework allows decoupling of each mode to \emph{arbitrary order} when noise leads to imperfect evolution of the composite system.  We first present a generic description of the method and then demonstrate its application in the context of Molmer-Sorensen (MS) gates for trapped-ion qubit pairs embedded in a linear chain~\cite{MUSIQC}, where residual coupling of ion internal states to multiple modes of motion leads to reduced gate fidelity.  This method complements existing optimal-control techniques~\cite{Kim2009,Choi2014}, but reduces technical complexity in gate implementation, permits suppression of noise in the drive, and nominally achieves the same decoupling operation in a shorter time.

We model the dynamical evolution of a compound system of $N$ qubits ($\mathcal{S}$) coupled to $M$ bosonic oscillator modes ($\mathcal{B}$) via the interaction Hamiltonian
\begin{equation}\label{eq:MShamMulti}
H_{\mathcal{SB}}(t)=i\hbar\sum_{\mu=1}^{N}\sigma_{\varsigma}^{\mu}
\sum_{k=1}^{M}\left(\gamma^{\mu}_{k}(t)a^{\dag}_{k}-\gamma^{\mu*}_{k}(t)a_{k}\right).
\end{equation}
Here $\sigma_{\varsigma}^{\mu}$ is a Pauli spin operator acting on the state of the $\mu$-th qubit, in a `direction' defined by the subscript $\varsigma\in\{x,y,z\}$, while $a^{\dag}_{k}$ ($a_{k}$) acts on $\mathcal{B}$ creating (annihilating) a single bosonic excitation of the $k$-th oscillator. Each of the functions $\gamma^{\mu}_{k}(t)$ has the form $\gamma^{\mu}_{k}(t)= f_{k}^{\mu}e^{i\delta_{k}t}r(t;\tau)$, where $\delta_{k}$ is the excitation frequency of the $k$-th oscillator and the coupling constant $f_{k}^{\mu}$ quantifies the strength of its interaction with the $\mu$-th qubit. The complex function $r(t;\tau)=\Theta[t]\Theta[\tau-t]e^{-i\phi(t)}$ represents a controlled temporal modulation of the coupling phase $\phi(t)$ implemented over an interval $t\in[0,\tau]$, during which the interaction Hamiltonian (\ref{eq:MShamMulti}) is effectively `switched on', generating the unitary operation
\begin{equation}\label{eq:totprop}
U(\tau)=\exp\left\{\sum_{\mu=1}^{N}\sigma_{\varsigma}^{\mu}B_{\mu}(\tau) +i\sum_{\mu,\nu=1}^{N}\varphi_{\mu\nu}(\tau)\sigma_{\varsigma}^{\mu}\sigma_{\varsigma}^{\nu}\right\}.
\end{equation}
For $N>1$, $\varphi_{\mu\nu}(\tau)\equiv\sum_{k}\text{Im}\int_{0}^{\tau}dt_{1}\int_{0}^{t_{1}}dt_{2}\gamma^{\mu}_{k}(t_{1})\gamma^{\nu*}_{k}(t_{2})$ represents an effective coupling between qubits $\mu$ and $\nu$ ($\mu\neq\nu$) that arises due to the state-dependent displacement of the oscillator system in phase-space, given by $B_{\mu}(\tau)\equiv\sum_{k=1}^{M}[f_{k}^{\mu}\alpha_{k}(\tau)a^{\dag}_{k}-f_{k}^{\mu*}\alpha_{k}^{*}(\tau)a_{k}]$, where $\alpha_{k}(\tau)\equiv\int_{0}^{\infty}dte^{i\delta_{k}t}r(t;\tau)$.

While the coupling interaction presented here is generic, it is frequently executed using a controlled periodic driving field.  In this context, the excitation frequency $\delta_{k}$ is realized via a detuning between the frequency of the driving field and a sideband associated with the $k$-th motional mode (formally due to a transformation to the interaction picture with respect to the free Hamiltonian).   The coupling strength $f_{k}^{\mu}$ is determined by the magnitude of the field, and $\phi(t)$ by its phase.

\begin{figure}[t]
\includegraphics[width=8.8cm]{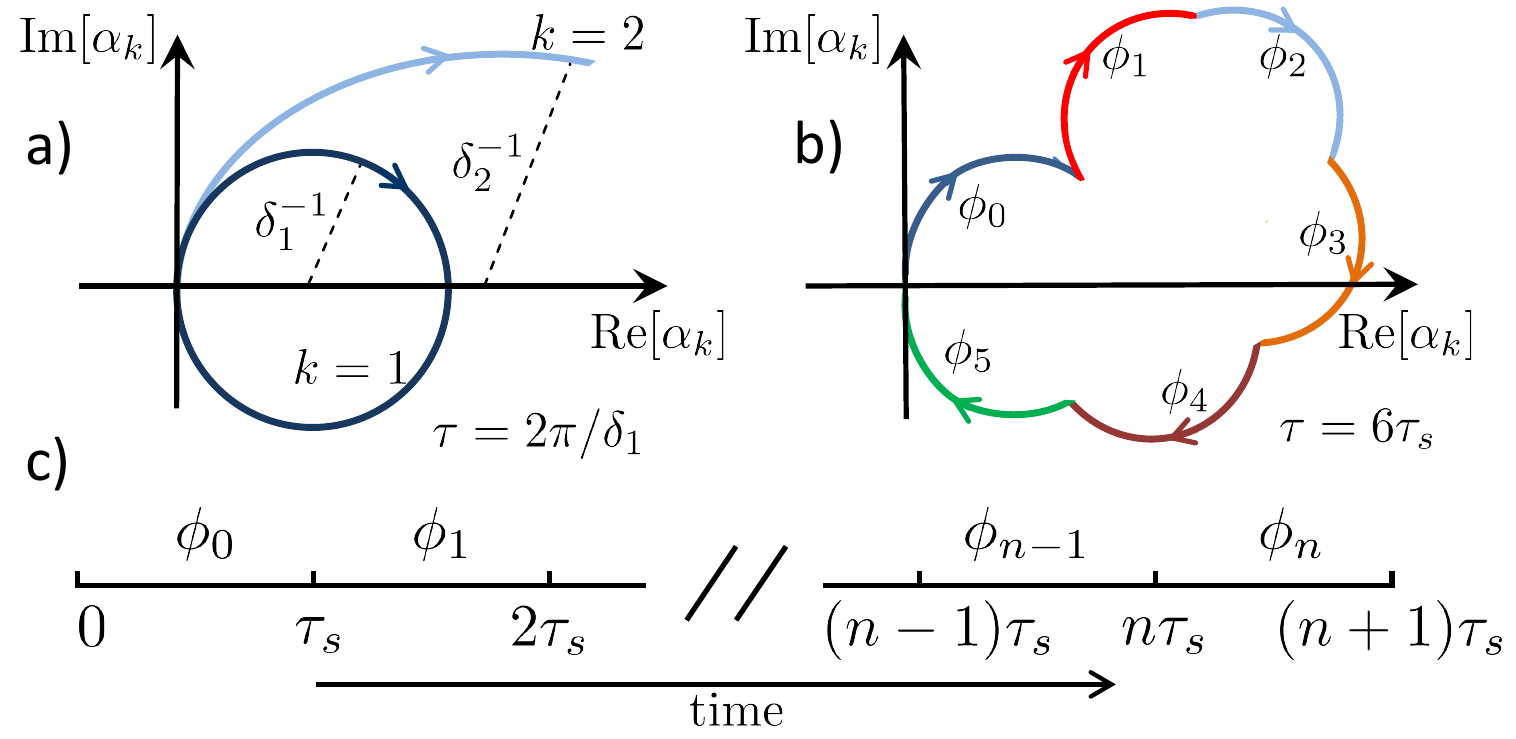}\\
\caption{\label{Fig:F1} a) Schematic plots of $\alpha_{k}(t)$, $0\leq t\leq\tau$, for two modes, $k=1,2$. Each path represents $N$ phase-space trajectories $\alpha^{\mu}_{k}(t)=f_{k}^{\mu}\alpha_{k}(t)$, for $\mu=1,...,N$. Here $\tau=2\pi/\delta_{1}$, so that the path labeled $k=1$ closes. b) An illustrative example of single closed path, generated by a six-interval, piecewise-constant phase modulation sequence. c) Timing schematic for piecewise-constant phase modulation sequence, comprising a total of $n$ instantaneous phase shifts. For the general $M$ mode decoupling sequence (\ref{eq:decouplingcond}) $n=2^{M}-1$.}
\end{figure}

Any residual qubit-oscillator entanglement at time $\tau$ will result in qubit decoherence and must, therefore, be suppressed in order to achieve a high-fidelity entangling operation. Complete qubit-oscillator decoupling occurs if
\begin{equation}\label{eq:deccond2}
\alpha_{k}(\tau)\equiv\int_{0}^{\infty}dte^{i\delta_{k}t}r(t;\tau)=0
\end{equation}
for $k=1,...,M$. Each of time-parameterized functions $\alpha_{k}(t)$, $0\leq t\leq\tau$, defines a set of $N$ phase space trajectories $\alpha^{\mu}_{k}(t)=f_{k}^{\mu}\alpha_{k}(t)$, for $\mu=1,...,N$, associated with the $k$-th oscillator mode (Fig. \ref{Fig:F1}). These trajectories vary in extent and orientation, according to the complex coupling constant $f_{k}^{\mu}$. However, by satisfying the condition (\ref{eq:deccond2}) all trajectories are closed at $t=\tau$.

Decoupling from any particular mode $k$ can be achieved by fixing the control phase at a constant value (which can be taken to be zero) and setting the total operation time and coupling-drive detuning such that $\delta_{k}\tau=2\pi j$, for $j\in\{1,2,...\}$ (Fig. \ref{Fig:F1}a). Simultaneous decoupling from any of the remaining modes is possible only if the associated detunings are commensurate with $\delta_{k}$. This can be difficult to engineer, even approximately, for more than one additional mode without resorting to undesirably long gate times \cite{Kim2009,Choi2014}.

Modulation of the relevant control, as demonstrated recently~\cite{Choi2014}, provides a path to simultaneously realizing the decoupling condition for multiple modes, and any of the parameters of the driving field may in principle be varied in time in order to achieve the necessary condition.  Unlike previous work, our method fixes both the frequency of the drive (hence $\delta_{k}$) as well as the coupling strength $f_{k}^{\mu}$ during the interaction period, and treats only the drive phase $\phi(t)$ as a tunable parameter.  In the following, we demonstrate that the freedom to modulate $\phi(t)$ via \emph{discrete shifts}, easily implemented using state-of-the-art digital frequency synthesis technology, may be used to impose a commensurate periodicity on the full set of phase-space paths so that they all close simultaneously at an (in principle) arbitrary time $\tau$.

The key to the method lies in the observation that any open phase space trajectory may be closed by repeating the sequence that produced it, \emph{with the appropriate overall phase shift}. To describe this process of `phase-compensated' concatenation mathematically, we define a family of \emph{nonlinear} operators $R_{\delta}$, parameterized by the real number $\delta$, that act to extend any phase modulation sequence $r(t;\tau')$, defined over $[0,\tau']$, to the interval $[0,2\tau']$ in the following way
\begin{equation}\label{eq:Rk}
R_{\delta}r(t;\tau')=r(t;\tau')+e^{-i(\delta\tau'-\pi)}r(t-\tau';\tau').
\end{equation}
The function $r(t;2\tau')\equiv R_{\delta_k}r(t;\tau')$ then describes a sequence for which $\alpha_{k}(2\tau')=0$.

Leveraging this operator, the qubit system may be simultaneously decoupled from all $M$ modes by implementing the piecewise-constant phase modulation sequence $r(t;2^{M}\tau_{s})\equiv R_{\delta_M}R_{\delta_{M-1}}...R_{\delta_1}r_{0}(t;\tau_{s})$, starting with the trivial `no-operation' base sequence $r_{0}(t;\tau_{s})=\Theta[t]\Theta[\tau_{s}-t]$, for which $\phi(t)\equiv0$ over an arbitrary interval $[0, \tau_{s}]$. This sequence, applied over discrete timesteps of duration $\tau_{s}$, indexed by $\ell$, may be written explicitly as
\begin{align}\label{eq:decouplingcond}
r(t;2^{M}\tau_{s})=\sum_{\ell=0}^{2^{M}-1}r_{0}(t-\ell\tau_{s};\tau_{s})e^{-i\phi_{\ell}}
\end{align}
where
\begin{align}\label{eq:phasevalue}
\phi_{\ell}=\sum_{j=0}^{q}\varepsilon_{j}(\ell)2^{j}\delta_{j+1}\tau_{s}-s(\ell)\pi
\end{align}
is the requisite phase value for the time interval $[\ell\tau_{s},(\ell+1)\tau_{s}]$ (see Fig. \ref{Fig:F1}). In this expression,  $\varepsilon_{q}(\ell)\varepsilon_{q-1}(\ell)...\varepsilon_{0}(\ell)$ is the binary representation of $\ell$ and $s(\ell)\equiv\sum_{j=0}^{q}\varepsilon_{j}(\ell)$ is its Hamming weight.  As an example, the binary representation of $\ell=6$ is $110$, so that $q=2$, $\varepsilon_{0}(\ell)=0$, $\varepsilon_{1}(\ell)=1$, $\varepsilon_{2}(\ell)=1$, and $s(\ell)$=2. The resulting `entangling phases' $\varphi_{\mu\nu}$ can then be calculated for these phase values (see \emph{Supplemental Material}), and the strength of the driving field adjusted to generate target values $\varphi^{(0)}_{\mu\nu}$, for $\mu,\nu=1,...,N$ and $\mu\neq\nu$.  

Ideally, the decoupling condition (\ref{eq:deccond2}) should also be met in the presence of time-domain variations in the coupling parameters defining phase-space trajectories. When only a single oscillator mode is coupled to the qubit system, discrete bivalued phase modulation in the form of simple binary ($\pm1$) concatenated dynamical decoupling (CDD) sequences~\cite{Khod2005,Khod2007} has been shown to suppress errors due to thermal dissipation \cite{Cen2005} and static detuning offsets \cite{Hayes2012}. In terms of our formalism, these sequences are generated by recursive application of the operator $R_{0}\equiv R_{\delta=0}$ to the simple no-operation base sequence $r_{0}(t;\tau_{s})$, with the proviso that the step time $\tau_{s}$ is chosen to coincide with the periodic evolution of the mode, i.e., $\delta_{k}\tau_{s}=2\pi j$, for $j=\{1,2...\}$.

The restriction on the allowable values of $\tau_{s}$ leads to proscription against direct application of CDD sequences for multiple modes.  We observe that by relaxing the binary-valued constraint on $\phi(t)$, phase compensated CDD sequences, targeting noise associated with \emph{particular modes}, may be realized.  Specifically, we consider noise that may be represented by a function $\beta_{k}(t)$ that modifies the spin-oscillator coupling via $\gamma^{\mu}_{k}(t)\rightarrow\gamma^{\mu}_{k}(t)\beta_{k}(t)$ (the subscript $k$ allows for the possibility of mode dependence.) We suppose that in the weak/slowly varying noise limit $\beta_{k}(t)$ may be approximated by a $p$-th order polynomial $\beta^{(p)}_{k}(t)=\sum_{j=0}^{p}\beta_{k,j}t^{j}$.  In the presence of such noise, the decoupling condition for the $k$-th mode \emph{to order} $p+1$ is
\begin{align}\label{eq:decouplingcondnoise}
\alpha^{(p+1)}_{k}(\tau)\equiv\int_{0}^{\infty}dte^{i\delta_{k}t}r(t;\tau)\beta^{(p)}_{k}(t)=0.
\end{align}
for $p\geq0$.

Mathematically, binary CDD sequences are essentially time-domain representations of finite iterations of the infinite Thue-Morse (TM) sequence~\cite{Allouche, Mauduit2001}. From the perspective of noise suppression, the most interesting property of the TM sequence is that the $(p+1)$-th iteration $r(t;2^{p+1}\tau_{s})\equiv R_{0}^{p+1}r_{0}(t;\tau_{s})$ is orthogonal to any $p$-th degree polynomial $\beta^{(p)}(t)=\sum_{j=0}^{p}\beta_{j}t^{j}$, i.e., $\int_{0}^{\infty}dt\;r(t;2^{p+1}\tau_{s})\beta^{(p)}(t)=0$ \cite{Richman2001}.  Using this insight, and despite the fact that we have relaxed the binary-value restriction on $\phi(t)$, it can be shown (see \emph{Supplemental Material}) that the phase-compensated TM/CDD sequence $r(t;(n+1)\tau_{s})\equiv R^{p+1}_{\delta_{k}}r_{0}(t;\tau_{s})=\sum_{\ell=0}^{n}r_{0}(t-\ell\tau_{s};\tau_{s})e^{-i\phi_{\ell}}
$, where $n=2^{p+1}-1$ and
$\phi_{\ell}=\ell\delta_{k}\tau_{s}-s(\ell)\pi$, will achieve $\alpha^{(p+1)}_{k}(2^{p+1}\tau_{s})=0$.

The general properties of concatenated control sequences may now be brought to bear in providing simultaneous high-order mode decoupling in the presence of time-varying coupling parameters.  To suppress noise across multiple modes, one may construct a phase modulation sequence of the general form $R_{\delta_{k_{q}}}...R_{\delta_{k_{2}}}R_{\delta_{k_{1}}}r_{0}(t;\tau_{s})$, where the \emph{order of error suppression} associated with mode $k_{i}$ is determined by the number of times $k_{i}$ appears in the sequence indices $(k_{1}, k_{2},...k_{q})$. For example, $R_{\delta_{3}}R_{\delta_{2}}R_{\delta_{3}}R_{\delta_{1}}r_{0}(t;\tau_{s})$ will close all trajectories associated with the first three modes, providing additional error suppression to second order for mode $k=3$. The order in which the operators are best applied will depend on the properties of the noise: in particular the extent to which the noise varies with $k$.  Vitally, in the presence of any such modulation protocol it remains possible to analytically calculate the entangling phase $\varphi_{\mu\nu}(\tau)$ for arbitrary qubit-pair $\mu-\nu$ and to adjust $f_{k}^{\mu(\nu)}$ appropriately.

\begin{figure}[t]
\includegraphics[width=8.8cm]{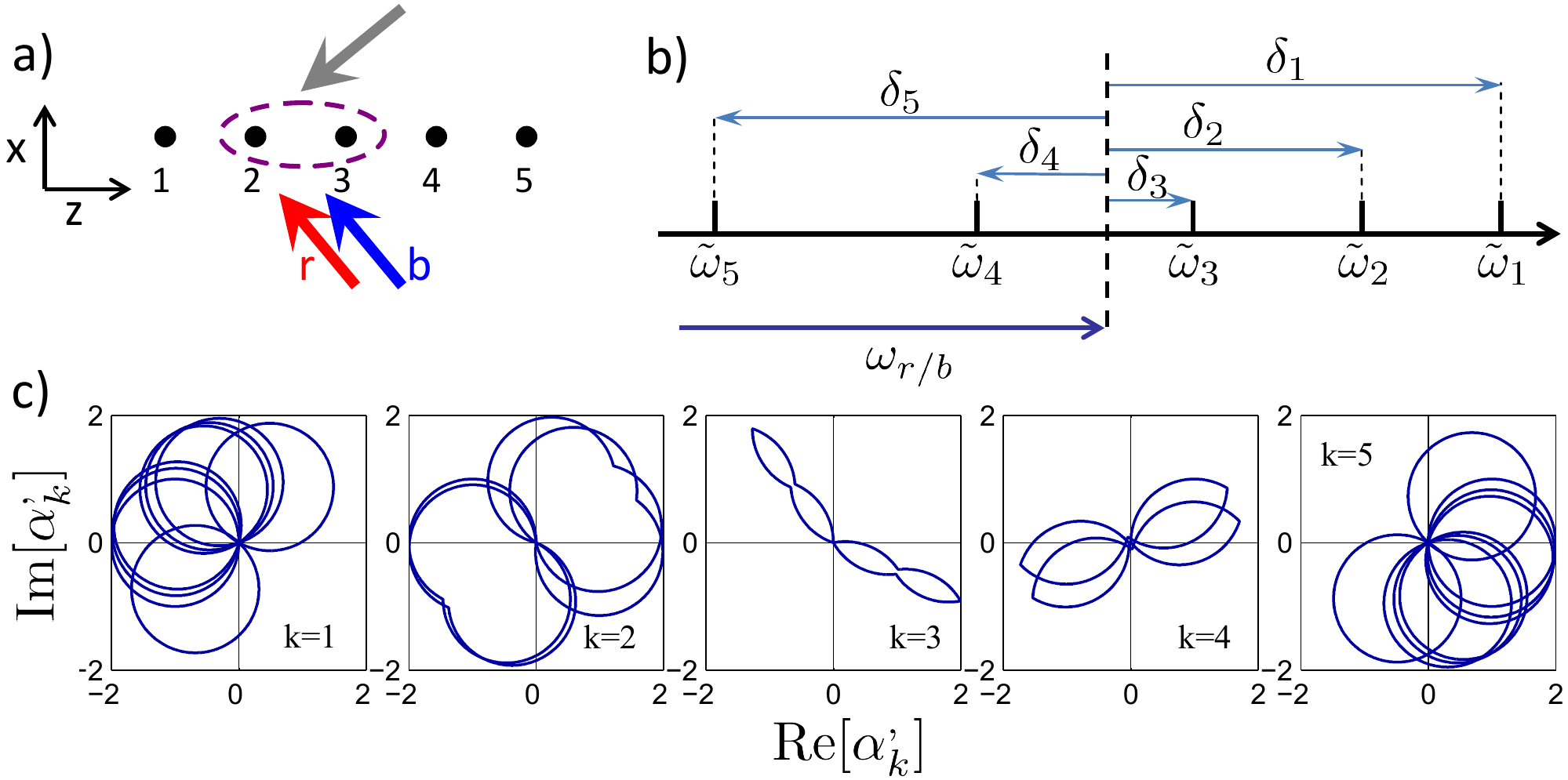}\\
\caption{\label{Fig:F2} a) Raman laser geometry for a MS gate applied to 2 ions in a 5 ion chain in which only transverse ($x$-direction) phonon modes are excited. The red (r) and blue (b) Raman fields have frequencies $\omega_{r/b}$ and phases $\phi_{r/b}$. b) Detuning diagram for 5 excited TP modes, $\tilde{\omega}_{k}=\omega_{0}\pm\omega_{k}$ ($+$ for $\omega_{b}$ and $-$ for $\omega_{r}$), where $\omega_{0}$ is the hyperfine qubit level splitting. c) Closed paths, $\alpha^{,}_{k}\equiv|\delta_{k}|\alpha_{k}(t)$, $0\leq t\leq\tau$, (normalized by $|\delta_{k}|$) for the detunings shown in b), generated by $7$ discrete phase shifts.}
\end{figure}

For concreteness, we now consider the task of entangling the internal states of a pair of adjacent trapped ions embedded in an $N>2$ ion chain.  Under certain simplifying assumptions, the effect of a state-dependent force generated by a bichromatic light field is well described by the Hamiltonian (\ref{eq:MShamMulti})~\cite{Sorensen1999, Lee2005, Brickman2010}. Effective spin-$1/2$ manifolds realized within the electronic states of each of the ions comprise the system of qubits $\mathcal{S}$, and the shared vibrational modes of the ions in a confining potential constitute the oscillator system $\mathcal{B}$. The qubit-oscillator coupling strength $f_{k}^{\mu}$ is proportional to the amplitude of the Raman fields (see \emph{Supplemental Material} for further details). The relevant control phase $\phi(t)$ is determined by the difference between the phases of the red and blue Raman fields, and can be varied without altering the spin-dependence of the entangling gate.  Most importantly, in this setting, $\phi(t)$ inherits all the flexibility and precision provided by modern laser control systems. In particular, the discrete phase shifts required for the basic decoupling sequence (\ref{eq:decouplingcond}) can be implemented quasi-instantaneously and with high accuracy using standard optical modulators driven by radiofrequency sources.

Fig \ref{Fig:F2}. shows closed mode trajectories representing the complete decoupling of a pair of $ ^{171}$Yb$^{+}$ hyperfine qubits from five excited transverse phonon modes~\cite{Zhu2006}, using phase-shifts derived from Eq.~\ref{eq:decouplingcond}. In this illustrative example, we set the laser frequencies so that two modes have commensurate detunings and choose $\tau_{s}=2\pi/\delta_{1,5}$ to match the period of the associated phase space evolution. In this way, a sequence of only $n=7$ phase shifts is required to decouple the qubits from all 5 modes, rather than the more general sequence of $n=31$ phase shifts. Assuming equivalent physical parameters to recent demonstrations of multimode decoupling using optimized amplitude modulation \cite{Choi2014}, the resulting phase-modulate gate has duration $\tau\sim140$ $\mu$s which compares favorably with the reported value of $\tau=190$ $\mu$s, while obviating considerations of nonlinear amplitude responses in optical modulators and rf amplifiers. Faster gate times may be achieved, at the expense of a greater number of phase shifts, by allowing $\tau_{s}$ to vary arbitrarily.

We also demonstrate the effectiveness of phase modulation sequences in suppressing decoupling errors induced by laser amplitude noise, a prominent time-dependent gate error.  We assume that a pair of qubits to be entangled is initially in an `evenly weighted' separable pure state, such as $|11\rangle_{z}=(|00\rangle_{x}+|11\rangle_{x}-|01\rangle_{x}-|10\rangle_{x})/2$ (the subscripts indicate the particular eigenbasis) and then quantify the extent of residual entanglement between the internal and vibrational ion states by calculating the linear entropy or purity loss $\bar{P}\equiv1-\text{Tr}[\rho_{\mathcal{S}}^{2}(\tau)]$, where $\rho_{\mathcal{S}}(\tau)$ is the final qubit state obtained by tracing out the vibrational degrees of freedom \cite{Tsallis1988,Audretsch2007}.

\begin{figure}[t]
\includegraphics[width=8.5cm]{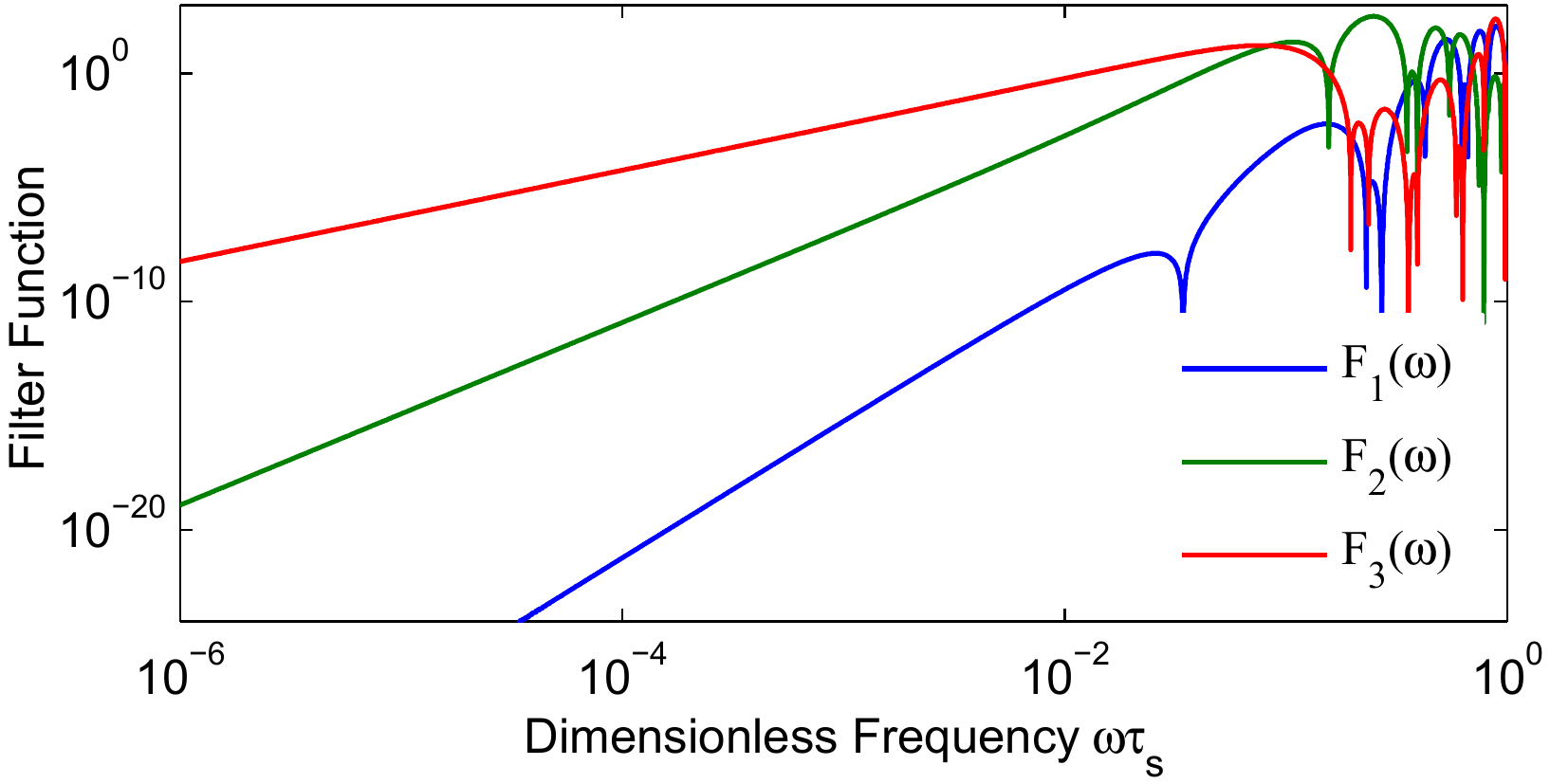}\\
\caption{\label{Fig:F3} Laser amplitude noise filter functions for 3 excited oscillator modes $k=1,2,3$. The effect of the noise on the coupling to each mode is suppressed to third order for $k=1$ ($F_{1}(\omega)\propto\omega^{3}$ as $\omega\rightarrow0$), second order for $k=2$ ($F_{2}(\omega)\propto\omega^{2}$), and first order for $k=3$ ($F_{3}(\omega)\propto\omega$) .}
\end{figure}

The laser amplitude instability is modeled as a time-dependent contribution to the Rabi rate $\Omega(t)=\Omega_{0}+\Omega_{e}(t)$, where $\Omega_{0}$ is the ideal value and $\Omega_{e}(t)$ represents the noise (we assume the Rabi rate is the same for the two adjacent ions). As $\Omega_{e}(t)$ is a stochastic process, we average over the ensemble of noise realizations ($\mathbb{E}[...]$) to obtain our final metric (see \emph{Supplemental Material} for a derivation)
\begin{equation}\label{eq:ensavpurity}
\mathbb{E}[\bar{P}]\approx\frac{1}{8\pi}\int_{-\infty}^{\infty}d\omega S_{\Omega_{e}}(\omega)F(\omega)
\end{equation}
Here, $S_{\Omega_{e}}(\omega)$ represents the power spectral density of amplitude fluctuations as a function of frequency $\omega$.  This expression assumes the weak noise limit in which only the first-order effect of amplitude fluctuations is significant, and assumes that phase modulation results in complete decoupling in the \emph{absence} of noise.

The effect of the phase modulation is captured by the expression $F(\omega)\equiv\sum_{k}D_{k}F_{k}(\omega)$, with $D_{k}$ a constant.  The `modal filter function' $F_{k}(\omega)\equiv|\int_{0}^{\infty}dte^{i(\omega+\delta_{k})t}r(t;\tau)|^2$ provides a frequency-domain representation of the phase modulation sequence $r(t;\tau)$, following insights presented in~\cite{KurizkiPRL2001, Green2013}.  It captures the effectiveness of the phase-modulation protocol in suppressing laser amplitude noise for different modes; a higher ``slope'' on a log-log plot indicates higher order ($p$) suppression of time-dependent fluctuations.  This general filter-function formalism for capturing the noise-suppressing properties of complex control protocols is validated by a range of single-qubit experiments~\cite{BiercukNature2009, SoareFiltering}, and has been extended here to qubit-oscillator and qubit-qubit interactions.

In Fig. \ref{Fig:F3} we plot $F_{k}(\omega)$ calculated for specific, but arbitrarily chosen, orders associated with each of three modes, $k=1,2,3$.  By increasing the level of concatenation for specific modes we are able to improve qubit-oscillator decoupling through the suppression of low-frequency amplitude fluctuations \emph{while simultaneously ensuring all modes are efficiently decoupled}.  In general, $D_{k}$ depends on the initial qubit state $|\phi_{0}\rangle$ and the effective temperature, in addition to the frequency of the $k$-th mode $\omega_{k}$. Here, where we consider only the collective zero temperature limit and the particular initial state $|\phi_{0}\rangle=|11\rangle_{z}$.

In summary, we have presented a unified framework for improving entangling gate fidelity in qubit-oscillator systems by ensuring efficient decoupling of intermediary bosonic modes.  Our framework permits simultaneous decoupling of multiple oscillator modes using only sequences of discrete phase shifts, and may also be combined with concatenation procedures to include robustness against time-dependent control noise.  This approach complements and provides substantial benefits relative to existing techniques, leveraging the technical simplicity of phase modulation in RF electronics.  It mitigates the need to account for nonlinearities in the response of modulating hardware (such as acousto-optic modulators for laser beams) or variations in residual light shifts endemic to Raman-mediated trapped-ion gates.  The relevant phase-shift sequences may be calculated simply using an intuitive framework, and our explicit calculations demonstrate that the gate can be implemented more rapidly than competing techniques.  This approach also builds in considerable freedom in the choice of step size $\tau_{s}$ and the order in which particular phase-space paths are closed.  Overall we hope that this phase-modulation approach to improving entangling gate fidelities will prove useful in a range of quantum information settings across different qubit-oscillator systems. We are also excited by the possibility that similar approaches may be employed to effectively modify the entangling phase between different qubits in a multipartite system, or be employed to improve the performance of quantum-enhanced sensors~\cite{Bowen2013}.  

\begin{acknowledgments}
\textit{Acknowledgements:}  The authors thank L. Viola, S.-W. Lee, G. Paz-Silva, and K. R. Brown  for useful discussions on dynamical decoupling as well as C. Monroe and J. Kim for useful discussions of motional mode structure and decoupling in linear ion traps.  This work partially supported by the Australian Research Council Centre of Excellence for Engineered Quantum Systems CE110001013, the Office of the Director of National Intelligence (ODNI), Intelligence Advanced Research Projects Activity (IARPA), through the Army Research Office, the Lockheed Martin Corporation, and a private grant from Hugh and Anne Harley.  

\end{acknowledgments}


\appendix
\subsection{Phase compensated TM sequences}
The $\ell$-th element of the infinite Thue-Morse (TM) sequence $(\mathfrak{T}_{\ell})_{\ell\geq0}$, defined over the two letter alphabet $\{-1, 1\}$, is given by $\mathfrak{T}_{\ell}=(-1)^{s(\ell)}$, where ${s(\ell)}$ is the Hamming weight of the binary representation of the integer $\ell$. Successive finite iterations of the sequence may be generated in the time domain through repeated application of the operator $R_{0}$ to the function $r_{0}(t;\tau_{s})$ (see main text for definitions). The $(p+1)$-th iteration $r(t;2^{p+1}\tau_{s})\equiv R_{0}^{p+1}r_{0}(t;\tau_{s})$ is orthogonal to any $p$-th degree polynomial $\beta^{(p)}(t)=\sum_{j=0}^{p}\beta_{j}t^{j}$, i.e., $\int_{0}^{\infty}dt\;r(t;2^{p+1}\tau_{s})\beta^{(p)}(t)=0$. This is a special case of the more general result that
\begin{align}\label{eq:SMdecouplingcondnoise}
\int_{0}^{\infty}dte^{i\delta t}r(t;2^{p+1}\tau_{s})\beta^{(p)}(t)=0
\end{align}
where $r(t;2^{p+1}\tau_{s})\equiv R^{p+1}_{\delta}r_{0}(t;\tau_{s})$, which we prove by induction.

Showing the result to be true for $p=0$ is simply a matter of direct substitution into (\ref{eq:SMdecouplingcondnoise}). Now assume (\ref{eq:SMdecouplingcondnoise}) holds for $p=q$, where $q$ is an arbitrary positive integer. In that case, it must hold for the $q$-th degree polynomial $\beta^{(q)}(t)\equiv\beta^{(q+1)}(t)-\beta^{(q+1)}(t+2^{q+1}\tau_{s})$. That is,
\begin{align}\label{eq:SMdecouplingcondnoise3}
&\int_{0}^{\infty}dte^{i\delta t}r(t;2^{q+1}\tau_{s})\beta^{(q+1)}(t)\notag\\
&-\int_{0}^{\infty}dte^{i\delta t}r(t;2^{q+1}\tau_{s})\beta^{(q+1)}(t+2^{q+1}\tau_{s})=0
\end{align}
Making the substitution $t\rightarrow t+2^{q+1}\tau_{s}$ in the second integral, we find that we can write
\begin{align}\label{eq:SMdecouplingcondnoise4}
\int_{0}^{\infty}dte^{i\delta t}r(t;2^{q+2}\tau_{s})\beta^{(q+1)}(t)=0
\end{align}
where $r(t;2^{q+2}\tau_{s})\equiv R^{q+2}_{\delta}r_{0}(t;\tau_{s})$. Thus, if (\ref{eq:SMdecouplingcondnoise}) holds for $q$, it also holds for $q+1$ and the proof is complete.

\subsection*{Entangling phases for a piecewise-constant phase-modulation sequence}
A piecewise-constant phase-modulation sequence defined over $n+1$ subintervals $\ell=0,...,n$, each of equal duration $\tau_{s}$, is described by the function
 \begin{align}\label{eq:SMpcseq}
r(t;(n+1)\tau_{s})=\sum_{\ell=0}^{2^{M}-1}r_{0}(t-\ell\tau_{s};\tau_{s})e^{-i\phi_{\ell}}.
\end{align}
Substituting this into the expression for the `entangling phase'
$\varphi_{\mu\nu}(\tau)=\sum_{k}\text{Im}\int_{0}^{\tau}dt_{1}\int_{0}^{t_{1}}dt_{2}\gamma^{\mu}_{k}(t_{1})\gamma^{\nu*}_{k}(t_{2})$, where $\gamma^{\mu}_{k}(t)= f_{k}^{\mu}e^{i\delta_{k}t}r(t;\tau)$, we find that
\begin{equation}\label{eq:SMentphase}
\varphi_{\mu\nu}(\tau)=\sum_{k}\frac{1}{\delta_{k}^2}
\left\{2(1-\cos{\delta_{k}\tau_{s}})A_{k}^{\mu\nu}+B_{k}^{\mu\nu}\right\}
\end{equation}
with
\begin{align}\label{eq:SMAB}
A_{k}^{\mu\nu}&\equiv\sum_{\ell=1}^{n}\sum_{\ell'<\ell}
\text{Im}\left\{f_{k}^{\mu}f_{k}^{\nu*}e^{i[(\ell-\ell')\delta_{k}\tau_{s}-\phi_{\ell}+\phi_{\ell'}]}\right\}
\end{align}
and
\begin{align}
B_{k}^{\mu\nu}&\equiv(n+1)\text{Im}\left\{f_{k}^{\mu}f_{k}^{\nu*}\left(i\delta_{k}\tau_{s}
-e^{i\delta_{k}\tau_{s}}+1\right)\right\}.
\end{align}

\subsection*{The Molmer Sorensen gate}
To implement the MS gate, the target ions are simultaneously illuminated by two off-resonant Raman laser fields with beat notes symmetrically detuned from motional sidebands on the qubit carrier frequency $\omega_{0}$ such that $\delta_{k}=\omega_{b}-(\omega_{0}+\omega_{k})=(\omega_{0}-\omega_{k})-\omega_{r}$. Here $\omega_{k}$ denotes the frequency of the $k$-th motional mode and the subscripts $b/r$ denote the blue ($b$) and red ($r$) Raman sidebands. This bichromatic field creates state-dependent forces on the illuminated ions that generate an effective interaction between the otherwise independent internal ion states \cite{Sorensen1999, Lee2005, Brickman2010}. The qubit-oscillator coupling strength $f_{k}^{\mu}=-i\Omega_{\mu}\eta_{\mu,k}/2$ is determined by the Rabi frequency $\Omega_{\mu}$, which is itself proportional to the common amplitude of the Raman fields, and by the the dimensionless Lambe-Dicke parameter $\eta_{\mu,k}$ \cite{Wineland1998}.

The spin dependence of the gate (i.e., the subscript $\varsigma$ in Eq. (\ref{eq:MShamMulti})) is decided by sum $\phi_{s}(t)=[\phi_{b}(t)+\phi_{r}(t)]/2$ of the phases of the red and blue Raman fields, while the control phase $\phi(t)$ (usually called the `motional phase') is given by their difference $\phi(t)=[\phi_{b}(t)-\phi_{r}(t)]/2$. By ensuring that the red and blue phases always have opposite sign, the control phase $\phi(t)\equiv\phi_{b}(t)=-\phi_{r}(t)$ and can be varied without altering the spin dependence (here $\varsigma=x$).

Assuming a common Rabi rate $\Omega=\Omega_{1}=\Omega_{2}$, the time-evolution operator for the two-qubit MS gate with piece-constant phase-modulation is
\begin{equation}\label{eq:totprop}
U(\tau)=\exp\left\{\sum_{\mu=1}^{2}\sigma_{x}^{\mu}B_{\mu}(\tau) +2i\varphi_{12}(\tau)\sigma_{x}^{1}\sigma_{x}^{2}\right\}
\end{equation}
where $B_{\mu}(\tau)\equiv\sum_{k,\ell}[\alpha_{k,\ell}^{\mu}(\tau)a^{\dag}_{k}-\alpha_{k,\ell}^{\mu*}(\tau)a_{k}]$, where
\begin{equation}
\alpha_{k,\ell}^{\mu}(\tau)\equiv-\frac{\Omega\eta_{\mu,k}}{2\delta_{k}}
e^{i(\delta_{k}\tau_{s}-\phi_{\ell})}\left(e^{i\delta_{k}\tau_{s}}-1\right).
\end{equation}
The entangling phase $\varphi_{12}(\tau)$ is given by equation (\ref{eq:SMentphase}), where now
\begin{align}\label{eq:SMAB}
A_{k}^{12}=\frac{\Omega^2\eta_{k,1}\eta_{k,2}}{4}\sum_{\ell=1}^{n}\sum_{\ell'<\ell}
\sin[(\ell-\ell')\delta_{k}\tau_{s}-\phi_{\ell}+\phi_{\ell'}]
\end{align}
and
\begin{align}
B_{k}^{12}=(n+1)\frac{\Omega^2\eta_{k,1}\eta_{k,2}}{4}\left(\delta_{k}\tau_{s}
-\sin{\delta_{k}\tau_{s}}\right).
\end{align}

\subsection{Example phase modulation sequence}

The table below lists the eight phase values $\phi_{\ell}$, for $\ell=0,...,7$, calculated using Eq.~\ref{eq:decouplingcond}, for the illustrative example described in the text. Execution of this sequence results in the complete decoupling of a pair of $ ^{171}$Yb$^{+}$ hyperfine qubits from five excited transverse phonon (TP) modes. The associated detunings are $\delta_{k}$$=$$2\pi\times$$\{59.77\text{kHz},$ $40.26\text{kHz},$ $11.06\text{kHz},$ $-20.07\text{kHz},$ $-59.77\text{kHz}\}$. The laser frequencies have been chosen such that modes $k=1,5$ have equal detunings and $\tau_{s}=2\pi/\delta_{1,5}$ matches the period of the associated phase space evolution.
\begin{table}[h]\label{tble:SM1}
\begin{tabular}{c|c}
\hline\hline
  \text{Interval} & \text{Phase}\\
  $\ell$ & $\phi_{\ell}$\\
\hline
  0 & 0 \\
  1 & $\delta_{1}\tau_{s}-\pi$ $=\pi$ \\
  2 & $2\delta_{2}\tau_{s}-\pi$ $\simeq1.694\pi$\\
  3 & $\delta_{1}\tau_{s}+2\delta_{2}\tau_{s}-2\pi$ $\simeq1.694\pi$\\\
  4 & $4\delta_{3}\tau_{s}-\pi\simeq0.4803\pi$\\
  5 & $\delta_{1}\tau_{s}+4\delta_{3}\tau_{s}-2\pi\simeq1.4803\pi$\\
  6 & $2\delta_{2}\tau_{s}+4\delta_{3}\tau_{s}-2\pi\simeq2.175\pi$\\
  7 & $\delta_{1}\tau_{s}+2\delta_{2}\tau_{s}+4\delta_{3}\tau_{s}-3\pi\simeq3.175\pi$\\
  \hline
  \end{tabular}
\end{table}

\subsection{Ensemble average purity loss}
The MS time-evolution operator (in the $\varsigma=x$ basis) may be written as $U(\tau)=U_{1}(\tau)U_{2}(\tau)$, where
\begin{equation}\label{eq:SMprop1}
U_{1}(\tau)=\exp\{\sum_{\mu}\sigma_{x}^{\mu}\sum_{k}[\alpha_{k}^{\mu}(\tau)a_{k}^{\dag}-\alpha_{k}^{\mu*}(\tau)a_{k}]\}
\end{equation}
and $U_{2}(\tau)=e^{i\varphi(\tau)\sigma^{1}_{x}\sigma_{x}^{2}}$. Here
\begin{equation}\label{eq:SMalpha}
\alpha_{k}^{\mu}(\tau)=\frac{-i\eta_{k}^{\mu}}{2}\int_{0}^{\infty}dt\Omega(t)e^{i\delta_{k}t}r(t;\tau)
\end{equation}
and
\begin{align}\label{eq:SMvarphi}
\varphi(\tau)&=\frac{1}{2}\sum_{k}\eta_{k}^{\mu}\eta_{k}^{\nu}
\int_{0}^{\tau}dt_{1}\Omega(t_{1})\int_{0}^{t_{1}}dt_{2}
\Omega(t_{2})\notag\\
&\hspace{1.5cm}\times\sin[\delta_{k}(t_{1}-t_{2})-(\phi(t_{1})-\phi(t_{2}))]
\end{align}
We assume that the combined qubit-oscillator system is in a separable initial state $|\psi_{0}\rangle\langle\psi_{0}|\otimes\rho_{\mathcal{B}}$, where
\begin{equation}\label{eq:SMinstate}
|\psi_{0}\rangle=\sum_{ij}c_{ij}|ij\rangle_{x}
\end{equation}
is a pure two-qubit state (expanded in the $x$ basis.). The final state is $\rho_{\mathcal{S}}(\tau)=U_{2}(\tau)\tilde{\rho}_{\mathcal{S}}(\tau) U_{2}^{\dag}(\tau)$,
where we've defined
\begin{align} \tilde{\rho}_{\mathcal{S}}(\tau)\equiv\text{Tr}_{\mathcal{B}}\left[U_{1}(\tau)|\psi_{0}\rangle\langle\psi_{0}|
\otimes\rho_{\mathcal{B}}U_{1}^{\dag}(\tau)\right]
\end{align}
Using the cyclic property of the trace, the purity loss may be written as $\bar{P}=1-\text{Tr}[\tilde{\rho}_{\mathcal{S}}^{2}(\tau)]$. Substituting the expansion (\ref{eq:SMinstate}) and assuming the oscillator system is in a state $\rho_{\mathcal{B}}=\prod_{k}\rho_{k}$, where $\rho_{k}=e^{-\hbar\omega_{k}a_{k}^{\dag}a_{k}/(k_{B}T_{k})}/\text{Tr}_{k}[e^{-\hbar\omega_{k}a_{k}^{\dag}a_{k}/(k_{B}T_{k})}]$ ($T_{k}$ is the effective temperature of the $k$-th mode), we can write
\begin{equation}\label{eq:SMtden}
\tilde{\rho}_{\mathcal{S}}=\sum_{ijlm}c_{ij}c_{lm}^{*}|ij\rangle\langle lm| e^{i\{s_{m}s_{i}-s_{l}s_{j}\}\sum_{k}\text{Im}(\alpha_{k}^{1}\alpha_{k}^{2*})}e^{-\chi_{ijlm}}
\end{equation}
where $\chi_{ijlm}=\sum_{k}|\tilde{\alpha}_{k}|^2\coth{[\hbar\omega_{k}/(2k_{B}T_{k})]}/2$, $\tilde{\alpha}_{k}=(s_{i}-s_{l})\alpha_{k}^{1}(\tau)+(s_{j}-s_{m})\alpha_{k}^{2}(\tau)$ and $s_{i}=(-1)^{i}$ (the eigenvalue of $\sigma_{x}$ for the state $|i\rangle_{x}$). From this, we find that
\begin{equation}\label{eq:SM}
\bar{P}=1-\sum_{ijlm}|c_{ij}|^{2}|c_{lm}|^{2}e^{-2\chi_{ijlm}}
\end{equation}
We now write the noisy Rabi rate (which is assumed to be real) as $\Omega(t)=\Omega_{0}+\Omega_{e}(t)$, where the function $\Omega_{e}(t)$ describes fluctuations about the ideal value $\Omega_{0}$. Further, we consider only phase modulation sequences for which perfect decoupling would occur in the absence of noise, i.e., $\alpha_{k}^{\mu}(\tau)=0$, for $\mu=1,2$ and $k=1,...,M$, when $\Omega_{e}(t)=0$. In that case,
\begin{equation}\label{eq:SMalpha2}
\alpha_{k}^{\mu}(\tau)=\frac{-i\eta_{k}^{\mu}}{2}\int_{0}^{\infty}dt\Omega_{e}(t)e^{i\delta_{k}t}r(t;\tau)
\end{equation}
and the exponent $\chi_{ijlm}$ depends only on the noise $\Omega_{e}(t)$ and not on $\Omega_{0}$.

We average over an ensemble of noise realizations to obtain the ensemble average ($\mathbb{E}[...]$) purity loss
\begin{equation}\label{eq:SM}
\mathbb{E}[\bar{P}]=1-\sum_{ijlm}|c_{ij}|^{2}|c_{lm}|^{2}\mathbb{E}[e^{-2\chi_{ijlm}}].
\end{equation}
Using the inequality $e^{x}\geq1+x$, we see that
\begin{align}\label{eq:SMpineq}
\mathbb{E}[\bar{P}]&\leq2\sum_{ijlm}|c_{ij}|^{2}|c_{lm}|^{2}\mathbb{E}[\chi_{ijlm}]
\end{align}
providing an upper bound on the purity loss. In the limit of weak noise the inequality approaches equality, giving an approximate weak noise expression for purity loss. To make the noise dependence explicit, we need to evaluate the quantities
\begin{align}\label{eq:SMchi1}
\mathbb{E}[\chi_{ijlm}]=&\frac{1}{2}\sum_{k}\mathbb{E}\left[|(s_{i}-s_{l})\alpha_{k}^{1}(\tau)
+(s_{j}-s_{m})\alpha_{k}^{2}(\tau)|^2\right]\notag\\
&\hspace{2cm}\times\coth{[\hbar\omega_{k}/(2k_{B}T_{k})]}
\end{align}
To do so, we first introduce $S_{\Omega_{e}}(\omega)$; the power spectral density of the Rabi rate fluctuations
\begin{equation}
\mathbb{E}[\Omega_{e}(t_{1})\Omega_{e}(t_{2})]
=\frac{1}{2\pi}\int_{-\infty}^{\infty}S_{\Omega_{e}}(\omega)e^{i\omega(t_{1}-t_{2})}.
\end{equation}
From (\ref{eq:SMpineq}) and (\ref{eq:SMchi1}), we then arrive at the weak noise approximation for the purity loss
\begin{equation}
\mathbb{E}[\bar{P}]\approx\frac{1}{8\pi}\int_{-\infty}^{\infty}S_{\Omega_{e}}(\omega)F(\omega)
\end{equation}
where $F(\omega)=\sum_{k}D_{k}F_{k}(\omega)$
\begin{align}
D_{k}&=\sum_{ijlm}|c_{ij}|^2|c_{lm}|^2
\left|(s_{i}-s_{l})\eta_{k}^{1}+(s_{j}-s_{m})\eta_{k}^{2}\right|^2\notag\\
&\hspace{3cm}\times\coth{[\hbar\omega_{k}/(2k_{B}T_{k})]}
\end{align}
and $F_{k}(\omega)=\left|\int_{0}^{\infty}e^{i(\omega+\delta_{k})t}r(t;\tau)\right|^2$.

\newpage

\end{document}